\documentstyle[aps,prl,epsf]{revtex}
\begin{document}
\draft
\twocolumn[\columnwidth\textwidth\csname@twocolumnfalse\endcsname
\title{Backbending in $^{50}$Cr}
\author{G. Mart\'{\i}nez-Pinedo$^{1)}$, A. Poves$^{1)}$,
L.M. Robledo$^{1)}$ \\ E. Caurier$^{2)}$,
F. Nowacki$^{2)}$, J. Retamosa$^{2)}$ and A. P. Zuker$^{1,2)}$}

\address{$^{1)}$ Departamento de F\'{\i}sica Te\'orica C-XI.
  Universidad Aut\'onoma de Madrid E--28049 Madrid, Spain}

\address{$^{2)}$ Division de Physique Theorique.  Centre de Recherches
  Nucleaires F-67037 Strasbourg Cedex-2 , France}

\preprint{FTUAM/96-18}

\maketitle

\begin{abstract}
  The collective yrast band and the high spin states of the nucleus
  $^{50}$Cr are studied using the spherical shell model and the HFB
  method. The two descriptions lead to nearly the same values for the
  relevant observables.  A first backbending is predicted at
  $I=10\hbar$ corresponding to a collective to non-collective
  transition.  At $I=16\hbar$ a second backbending occurs, associated
  to a configuration change that can also be interpreted as an
  spherical to triaxial transition.
\end{abstract}


\addvspace{5mm}
]



In a recent paper \cite{NOS} we have shown that large scale Shell
Model (SM) calculations with the realistic interaction KB3 predict the
same intrinsic state than mean field calculations (Cranked
Hartree-Fock-Bogoliubov) with the Gogny force for the ground state
rotational band of $^{48}$Cr. Thus, we have two complementary views 
of the problem. For the SM wave functions have the
proper quantum numbers (angular momentum and particle number) and
include correlations in the wave functions needed for a detailed
account of the observables, while the mean field results
provide us with a simpler understanding of the intrinsic state on top
of which the rotational ground state band is built. In this paper we
proceed in the same way for the nucleus $^{50}$Cr.
Experimentally~\cite{NDS}, it is known that up to $J=10$ the Yrast
states of $^{50}$Cr are members of a rotational band. The gamma ray
energy of the last known state $J=12^+$ to the $J=10^+$ 
indicate that $^{50}$Cr has started to backbend at $J=10$  this
situation being reminiscent of that  encountered in $^{48}$Cr.  The two
additional neutrons of the nucleus $^{50}$Cr will contribute to the
appearance of new structures when the angular momentum of the system
is increased.  As in the case of the Yrast band of $^{48}$Cr, where
the backbending was due to a deformed to spherical transition, we find
in $^{50}$Cr that the incipient backbending observed at $J=10$ is also
due to a shape transition from the prolate $J=10$ state to the nearly
spherical (actually slightly oblate) $J=12$ state. Contrary to what
happens in $^{48}$Cr the Yrast band of $^{50}$Cr undergoes a second
transition at $J=16$ from the slightly oblate regime to a more
deformed, triaxial regime that causes a second backbend. The behaviour
of the Yrast band of $^{50}$Cr can also be understood in terms of the
changes of the occupancies of the $1f_{7/2}$ orbitals of protons and
neutrons.

\vskip 0.5cm
{\bf Computational procedures.}
We shall briefly describe the computational procedures used
in this paper. A more detailed account can be found in~\cite{NOS}. In
the Spherical Shell Model (SM) $^{50}$Cr is described in a $0\hbar
\omega$ space, i.e. ten particles are allowed to occupy all the states
available in the $pf$ shell ($\sim 10^7$). The effective interaction
is the same used in the $^{48}$Cr calculation i.e., a minimally
modified version of the Kuo-Brown's G-matrix~\cite{MSSMint} denoted
KB3 in~\cite{report}. The single particle energies are taken from the
$^{41}$Ca experimental spectrum.  The effect of core polarization on
the quadrupole properties is taken into account by the use of
effective charges $q_\pi =1.5$, $q_\nu =0.5$.  These reasonable values
agree with the estimation of core effects in the quadrupole properties
carried out for $^{48}$Cr~\cite{NOS}.  The secular problem is solved
using the code ANTOINE~\cite{ANTOINE}, a very fast and efficient
implementation of the Lanczos method.

In the intrinsic frame calculations we have used the Self Consistent
Cranking Hartree-Fock-Bogoliubov method (HFB) with the density
dependent Gogny force~\cite{gogny}. The mean field intrinsic states $
|\phi_\omega \rangle $ have been expanded in a triaxial harmonic
oscillator basis $|n_xn_yn_z\rangle $ with different oscillator
lengths. Ten oscillator shells are included in this calculation in
order to ensure the convergence of the mean field results.  In our
calculation we use the DS1 parameters set of the Gogny
force~\cite{GognyPar}.  Without further changes, this force has proven
capable of describing successfully many phenomena, and in particular
high spin behaviour~\cite{CHFBcal}.

\begin{figure}
  \begin{center}
    \leavevmode
    \epsfxsize=8cm
    \epsffile{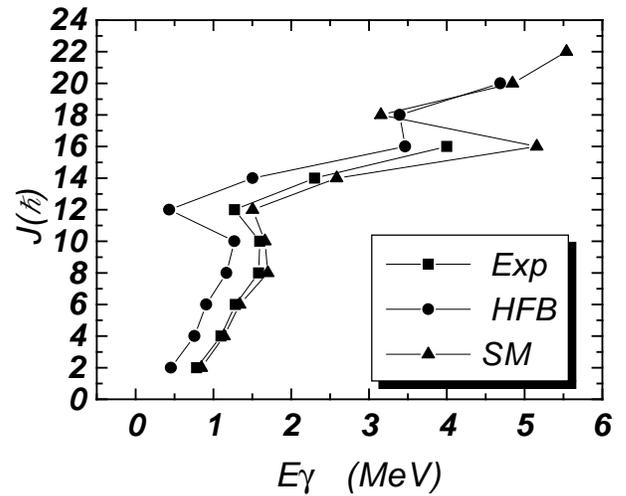}
  \end{center}
  \caption{Theoretical (triangles, SM; circles HFB) and
    experimental (squares) gamma ray energies versus the angular
    momentum $J$}
  \label{fig:gammaray}
\end{figure}

\vskip 0.5cm
{\bf The Yrast band.}
In fig. \ref{fig:gammaray} the SM, HFB and experimental gamma ray
energies $E_\gamma (J) = E(J) -E(J-2)$ are plotted as a function of
the angular momentum $J$ for the Yrast band. The SM results are very
close to the experimental data. The CHFB results follow the trend of
the SM ones but are shifted downwards in energy indicating a much
bigger static moment of inertia than in the SM and the experiment.
In~\cite{NOS} it was argued that this behaviour is the result of a
deficient treatment of pairing correlations by the HFB method in the
weak pairing regime. However, this deficiency does not affect to the
nature of the intrinsic state in a substantial way. The experimental
results from the Chalk-River Mc Master collaboration~\cite{cameron},
show preliminary evidence for Yrast states (14$^+$ and 16$^+$) at
excitation energies 9.9 MeV and 13.9 MeV. These results would confirm
the existence of a first backbending in $^{50}$Cr. Notice that the
corresponding E$_\gamma$'s, 2.3 MeV and 4.0 MeV fit reasonably well
into our predictions.

In both theoretical calculations a second backbending is predicted to
take place at $J=16 \hbar$.  Its origin will be discussed below in
terms of the changes observed in the quadrupole moment and of the
evolution of the occupancies of the relevant spherical orbits.

\begin{figure}
  \begin{center}
    \leavevmode
    \epsfxsize=8cm
    \epsffile{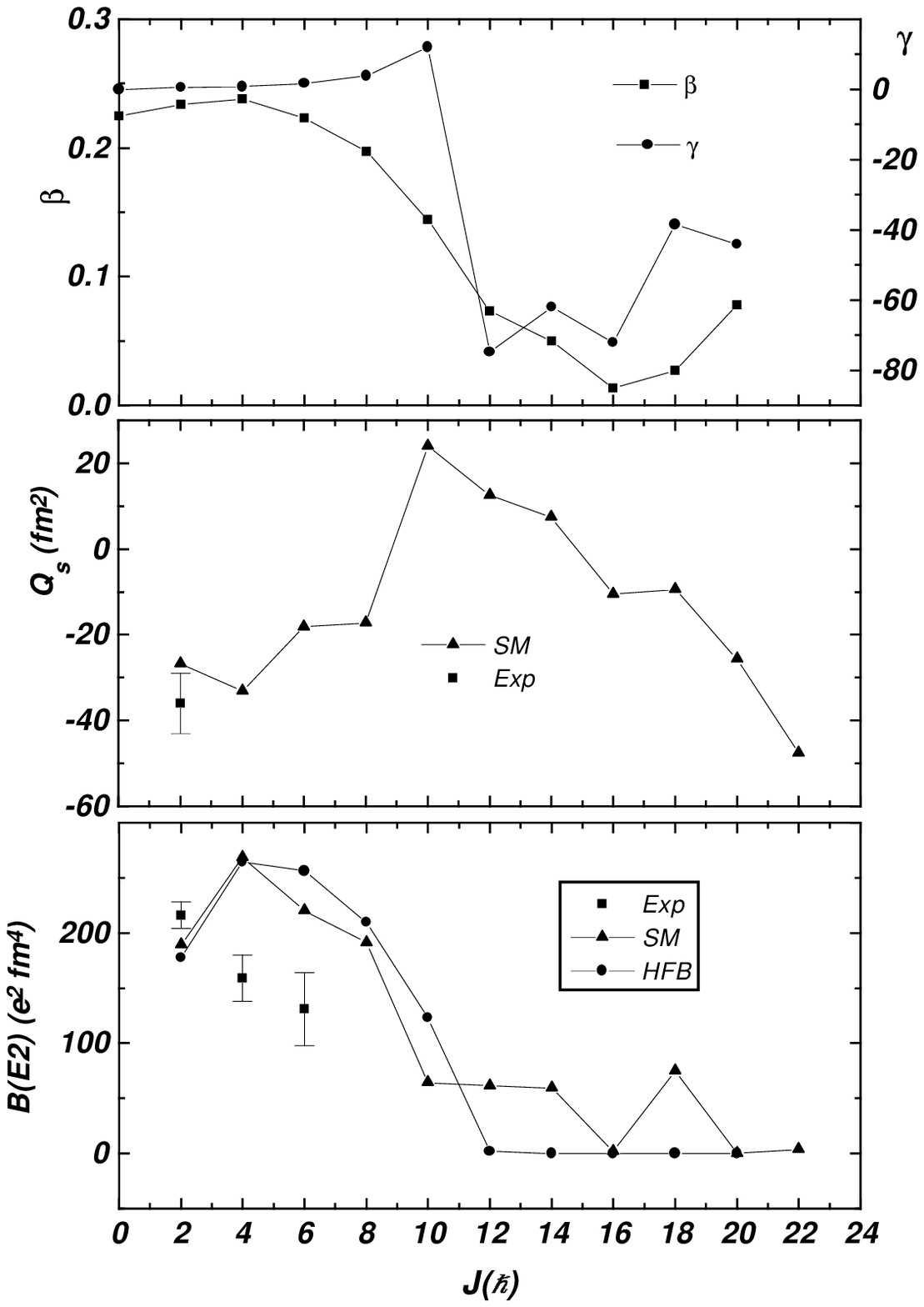}
  \end{center}
  \caption{Upper panel: $\beta$ and $\gamma$ deformations in the HFB
    calculation. Middel panel: Spectroscopic quadrupole moment $Q_s$
    computed in the SM approach. The experimental value
    $Q_s(2^+)$ is also given.  Lower panel: $B(E2,J\rightarrow J-2$
    transition probabilites versus $J$ computed in SM (triangles) 
    and HFB (circles) compared to the
    experimental data (squares)}
\label{fig:Quadrupole}
\end{figure}

The quadrupole properties obtained in both approaches are summarized
in Fig.~\ref{fig:Quadrupole}. In the upper panel the $\beta$ and
$\gamma$ deformation parameters obtained in the HFB calculation are
plotted versus the angular momentum. We can interpret these results as
defining three regions:

\begin{itemize}
\item[---] At low angular momentum $^{50}$Cr is an axially symmetric
  prolate nucleus ($\beta \sim 0.23$ and $\gamma \sim 0$)
  
\item[---] At $J=10\hbar$ the $\beta$ value drops to 0.15 but the
  system is still axially symmetric while at $J=12\hbar$ the system
  undergoes a shape transition to a weakly deformed oblate state with
  $\beta =0.08$ and $\gamma = -67$ degrees. This sudden change is
  responsible for the observed backbending at $J=10 \hbar$.
  
\item[---] At $J=16\hbar$ a new transition occurs that makes the
  system triaxial increasing at the same time the $\beta $ value.
  These changes of the quadrupole moment are responsible for the
  backbending seen at $J=16\hbar$.
\end{itemize}

In the middle panel the spectroscopic quadrupole moment of the SM
calculation is plotted as a function of $J$. $Q_s$ changes sign, from
negative to positive at $J=10$. This point has been recently
discussed by Zamick and coworkers~\cite{zamick1}. They made truncated
shell model calculations using also the interaction KB3. Their results
for the quadrupole moments at truncation level $t$=3 are qualitatively
equivalent to our full space results. Nevertheless, they interpret
these positive quadrupole moments as pertaining to a ``high K
prolate'' band~\cite{zamick2} while we rather rely in the HFB results
to conclude that these states are of oblate, non-collective character.
This transition is correlated with the backbending in the $E_\gamma$
plot.  The spectroscopic quadrupole moment and the intrinsic
deformations $\beta$ and $\gamma$ behave in a very similar way.
However, it is difficult to establish a close connection between them
in the high spin region where the $K$ labeling is probably
meaningless. In ref.~\cite{zamick1} it was pointed out that the very
large and negative values of $Q_s$ for the states of highest spins
correspond to those for the maximally aligned configurations. Our
results agree with this interpretation. Triaxiality comes in because
two or more of these configurations are simultaneously present in the
wave function of these yrast states.

In the lower panel of Fig.~\ref{fig:Quadrupole} the $B(E2)$
transition probabilities from the SM and HFB calculations are plotted
as a function of $J$ and compared with the available experimental
data. The HFB results are obtained from the intrinsic values of the
quadrupole operators $Q_{20}$ and $Q_{22}$ using an improved
rotational formula~\cite{EGWM}. Both results are very similar in the
whole range of angular momentum considered. The comparison with the
experiment data is good for the $B(E2,2^+\rightarrow0^+)$ but both
theoretical results overestimate the experimental ones at $J=4 $ and
$6 \hbar$.

\begin{figure}
  \begin{center}
    \leavevmode
    \epsfxsize=7cm
    \epsffile{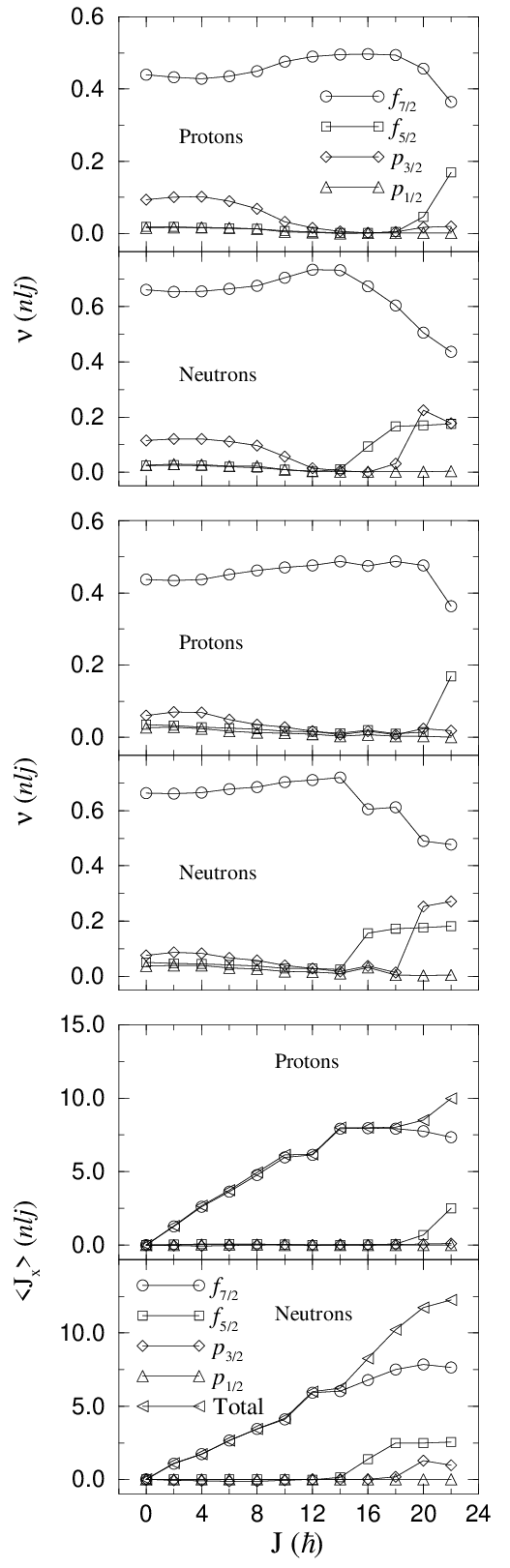}
  \end{center}
  \caption{Upper panel: ``Fractional shell occupancies'' $\nu (n,l)$
    computed in the HFB approach as a function of $J$. Middel panel:
    Same as before but for the SM. Lower panel: ``Shell contribution
    to $\langle J_x \rangle$'' $j_x (n,l)$ in HFB.}
  \label{fig:occ}
\end{figure}

In order to understand the behaviour of the previously considered
observables as a function of $J$ it is convenient to analyze the
structure of the intrinsic states in terms of the ``fractional shell
occupancies'' $\nu (n,l)$ and the ``shell contribution to $\langle J_x
\rangle$'' $j_x (n,l)$ defined in ref.~\cite{NOS}.  In
Figure~\ref{fig:occ} we have plotted the fractional occupancies of the
spherical orbits in the HFB solution (upper panel) and in the SM one
(middle panel). In addition the shell contributions to $\langle J_x
\rangle$ (which only make sense in the HFB calculation) are plotted
for the relevant spherical orbits in the lower panel. There are
striking similarities between the SM and HFB occupancies, with the HFB
results being slightly smoother than the SM ones. Let us examine these
numbers for the three regions sketched before.

Up to $J=8\hbar$ only the $1f_{7/2}$ orbit and its $\Delta j=2$
deformation partner $2p_{3/2}$ are appreciably occupied. This is in
agreement with the fact that up to that value of the angular momentum
the nucleus is an axially symmetric prolate rotor ($\beta \sim 0.23$)
as surmised in ref.~\cite{zrpc}. The situation is the same than in
$^{48}$Cr.  For $J=8\hbar$ the occupation of the $2p_{3/2}$ orbit
starts decreasing in favor of the $1f_{7/2}$, as the later is more
capable to create angular momentum. As a consequence the $\beta$
deformation gets reduced (see Fig.~\ref{fig:Quadrupole}) and the first
backbending sets in.

At $J=12 \hbar$ the $2p_{3/2} $ orbit is emptied both for protons and
neutrons eliminating the quadrupole coherence. In the lower panel we
can see that the contribution of the neutrons to the total angular
momentum, $J=12\hbar$, is $J=6\hbar$ and comes fully from the
$1f_{7/2}$ orbit. But, six units of angular momentum is the maximum
attainable by six identical particles in the $1f_{7/2}$ orbit,
therefore neutrons are fully aligned.  In order to build up fourteen
units of angular momentum the system chooses to align the four protons
in the $1f_{7/2}$ orbit to obtain $J=8\hbar$ (the other alternative is
to move neutrons to higher orbits).  $J=14\hbar$ is the maximum
angular momentum that can obtained with four protons and six neutrons
in the $1f_{7/2}$ orbit. Therefore, to reach $J=16\hbar$, the system
is forced to pump out particles of the low $K$ states of the
$1f_{7/2}$ orbit to high $K$ states of the $1f_{5/2}$ as seen in the
neutron occupancies plot. This leads to the second backbending because
the excitation energy of the 16$^+$ state is anomalously large. When
the $1f_{5/2}$ neutron shell is exhausted (at $J=18\hbar$) the
$2p_{3/2}$ neutron shell enters into the game again.  Placing
particles in the $2p_{3/2}$ increases again the quadrupole coherence
producing the increase of deformation observed in
Fig.~\ref{fig:Quadrupole}. At $J=22\hbar$ the band terminates as the
maximum angular momentum that can be built with four protons and six
neutrons in the $fp$ shell is $23\hbar$.

\begin{figure}
    \begin{center}
    \leavevmode
    \epsfxsize=8cm
    \epsffile{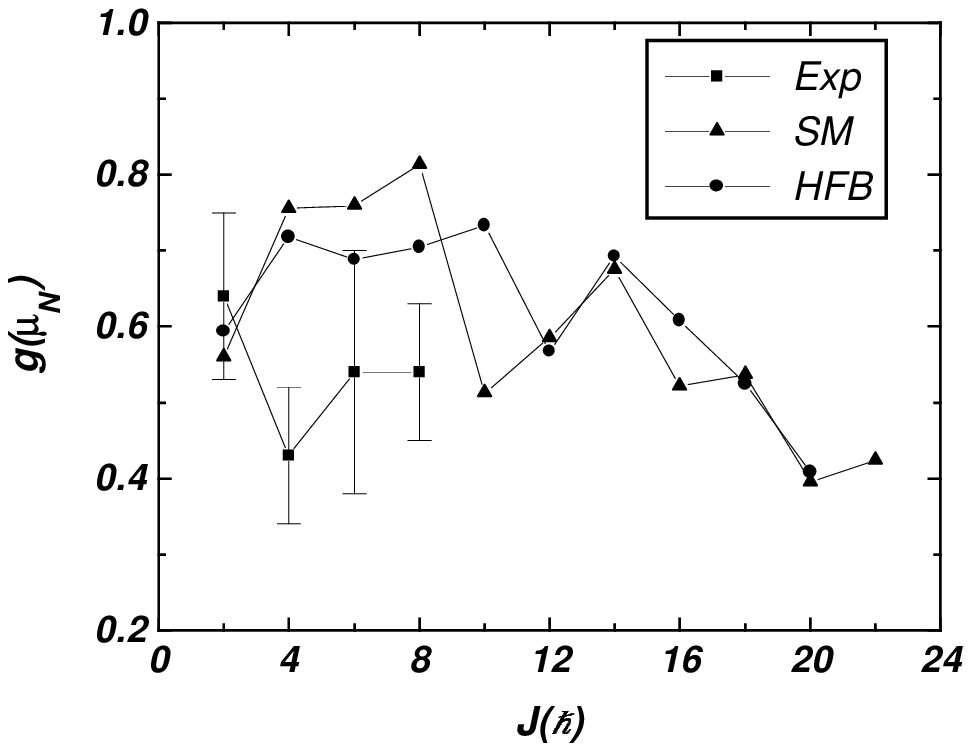}
  \end{center}
  \caption{Gyromagnetic factors: SM (triangles), HFB (circles) and
    experiment (squares).} 
\label{fig:mag}
\end{figure}

In figure \ref{fig:mag} we present our predictions for the
gyromagnetic factors compared with the experimental results of Pakou
{\it et al.}~\cite{Pakou94}. Once again the SM and HFB results are
nearly identical for the whole band but they do not agree with the
data except for the state $J=2$. The situation is somehow puzzling. We
know that $^{50}$Cr is not as good a rotor as $^{48}$Cr; its B(E2)'s
are smaller and the collectivity, as a function of the rotational
frequency, vanishes earlier. Our descriptions of $^{50}$Cr tend to
overshoot the experimental B(E2) values, indicating that possibly we
are obtaining a too large quadrupole collectivity. However, for the
gyromagnetic factors, the contrary seems to happen. We need to
increase the quadrupole correlations in order to approach the
rotational limit indicated by the experiment. It is difficult to move
simultaneously in both directions. We have tried another version of
the KB interaction called KB' in~\cite{pfcal} that has a smaller gap
between the $2p_{3/2}$ and the $1f_{7/2}$ orbits, thus favoring
deformation. As we feared, while the $g$ values get reduced
($g$(2$^+$)=0.53, $g$(4$^+$)=0.60, $g$(6$^+$)=0.62, $g$(8$^+$)=0.66)
and nearly agree with the experimental results of ref.~\cite{Pakou94},
the B(E2)'s increase a lot. For instance the 4$^+$ $\rightarrow$ 2$^+$
transition has B(E2)=415 e$^2$fm$^4$, compared to the experimental
value 159(21) e$^2$fm$^4$ or to the KB3 value 264 e$^2$fm$^4$.
Therefore we think that it will be extremely difficult to reproduce
simultaneously the experimental $g$ factors and the experimental
B(E2)'s.

\vskip 0.5cm
{\bf Further discussion of the high spin region.}
We have already shown that the double backbending in
$^{50}$Cr is related to the existence of three different zones in the
yrast line, and takes place in the two interphases. In the first of
those regions, up to $J=10\hbar$ the decay proceeds by the usual
sequence of $\Delta J=2$ enhanced E2 transitions. Here we shall study
the decay patterns in the other two zones.

\begin{figure}
  \begin{center}
    \leavevmode 
    \epsfxsize=8cm 
    \epsffile{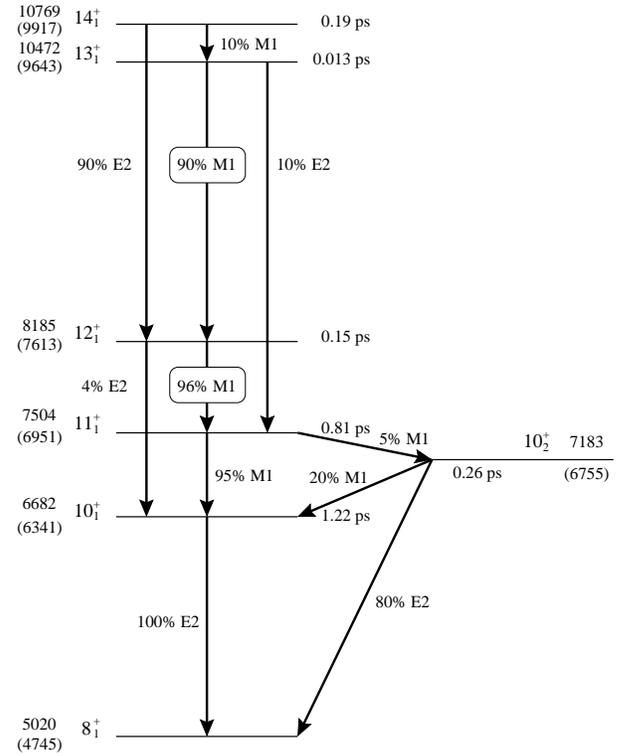}
  \end{center}  
  \caption{Decay scheme in the region of the first
    backbending. Energies in keV. In parenthesis the experimental
    values.} 
  \label{fig:decay}
\end{figure}

In figure \ref{fig:decay} we have plotted the yrast band as given by
the SM calculation, from the first backbending to the maximum $J$
attainable in the $(1f_{7/2})^{10}$ configuration. A new feature is
evident: beyond $J=10\hbar$ the yrast sequence shows up as $\Delta
J=1$. Therefore, the decay may become more complex, in particular M1
transitions are now allowed. And not only allowed but dominant due to
the hampering of the quadrupole collectivity in this zone.  In the
figure we can follow the decay subsequent to the feeding of the
$J=14\hbar$ state. Indeed the emitted gammas are mostly M1, only the
$14^+$ $\rightarrow$ $12^+$ is dominantly E2. The figure also presents
our predictions for the lifetimes and branching ratios using the
measured E$_{\gamma}$'s.  Experimental results from the Mc Master
Chalk River collaboration and from Legnaro~\cite{lenzi} are expected
to come out soon.

It is quite exciting to find out that the yrast decay at the first
backbending proceeds via magnetic transitions.

Let's now move higher in energy and angular momentum. If we assume
that the yrast band is fed at the yrast 18$^+$ state, the dominant
decay sequence will be:

\begin{displaymath}
18^+_1 \stackrel{E2}{\longrightarrow} 16^+_1
         \stackrel{E2}{\longrightarrow} 14^+_2
         \stackrel{E2}{\longrightarrow} 12^+_1
\end{displaymath}

If we choose the yrast 17$^+$ state as entry point; the decay sequence
will be:

\begin{displaymath}
17^+_1 \stackrel{E2}{\longrightarrow} 15^+_1 
         \stackrel{E2}{\longrightarrow} 13^+_2 
         \stackrel{M1}{\longrightarrow} 12^+_1  
\end{displaymath}

Hence, in the third zone we recover the $\Delta J=2$, E2 decay
pattern, in spite of the fact that the yrast sequence is now 14$^+$ -
15$^+$ - 16$^+$ - 17$^+$ - 18$^+$. The surprising thing is that the
decay bypasses the yrast 14$^+$ and 13$^+$ states, a striking feature
associated to the second backbending.

\vskip 0.5cm
In conclusion, we have shown that the spherical shell model and the
 Cranked Hartree Fock Bogolyubov method give similar -and
accurate- descriptions of $^{50}$Cr.  In particular, they
predict two backbendings in the E$_\gamma$ plot of the yrast band,
which are interpreted as due to two changes of regime: first from
collective prolate to non-collective oblate and second from
$(1f_{7/2})^{10}$-dominated configurations to aligned states with less
than ten particles in the $1f_{7/2}$ shell. A puzzling situation is
found in the collective, low energy part of the yrast band, where the
calculations fail to reproduce the experimental gyromagnetic factors.
Finally in our study of the decay sequence from the high spin yrast
states we have found out that the decay mode varies from one zone to
another. In the high spin regime it proceeds by stretched E2
transitions, at the first backbending by M1's and in the lower part by
collective E2's.  At the second backbending the decay sequence
bypasses the yrast line .

\vskip 0.5cm
We thank Dr. J. A. Cameron for making his results available to us
before publication.  This work has been partially supported by the
DGICyT, Spain under grants PB93-263 and PB91--0006, and by the
IN2P3-(France) CICYT (Spain) agreements. A. Z. is Iberdrola Visiting
Professor at the Universidad Aut\'onoma de Madrid.


\newpage

\end{document}